\documentclass[journal]{IEEEtranTIE}

\usepackage[utf8]{inputenc}     
\usepackage[T1]{fontenc}
\usepackage{graphicx}
\usepackage{amsmath}
\usepackage{amssymb}
\usepackage{epstopdf}
\usepackage{flushend}
\usepackage[hidelinks]{hyperref}
\usepackage{soul}           
\usepackage{subfigure}
\usepackage{float}
\usepackage{multicol}
\usepackage[most]{tcolorbox}
\usepackage{natbib}
\usepackage{multirow}
\usepackage[figuresright]{rotating}
\usepackage{placeins}

\newcommand{\chip}{{\,{chip}}}
\newcommand{\fm}{{\,{fm}}}
\newcommand{\duct}{{\,{duct}}}
\newcommand{\res}{{\,{res}}}
\newcommand{\lin}{{\,{line}}}
\newcommand{\reg}{{\,{reg}}}
\newcommand{\out}{{{out}}}
\newcommand{\atm}{{{atm}}}
\newcommand{\tot}{{{tot}}}
\newcommand{\Res}{\mathcal{R}} 
\newcommand{\Qkf}{\mathbf{Q_{\scriptscriptstyle KF}}}
\newcommand{\Rkf}{\mathbf{R_{\scriptscriptstyle KF}}}
\newcommand{\Qmpc}{\mathbf{W_y}}
\newcommand{\Rmpc}{\mathbf{W_u}}

\begin{document}

\title{\huge{Dynamic modeling and predictive control\\of a microfluidic system}}

\author{Jorge Vicente Mart\'inez, Edgar Ramirez-Laboreo, and~Pablo Calder\'on Gil
    \thanks{J. Vicente Mart\'inez and E. Ramirez-Laboreo are with the Departamento de Informatica e Ingenieria de Sistemas (DIIS) and the Instituto de Investigacion en Ingenieria de Aragon (I3A), Universidad de Zaragoza, 50018 Zaragoza, Spain (e-mail: j.vicente@unizar.es; ramirlab@unizar.es).}
    \thanks{P. Calder\'on Gil is with the Instituto Tecnológico de Arag\'on (ITAINNOVA), 50018 Zaragoza, Spain (e-mail: pcalderon@itainnova.es).}
    \thanks{\textcolor{red}{This is an author-approved English translation of the following work: Vicente Mart\'inez, J., Ram\'irez Laboreo, É., y Calder\'on Gil, P. (2024) ``Modelado din\'amico y control predictivo de un sistema microflu\'idico'', \textit{Revista Iberoamericana de Autom\'atica e Inform\'atica industrial}. DOI: 10.4995/riai.2024.19953. \textbf{Please cite the publisher's version}. For the publisher's version and full citation details see: \protect\href{https://doi.org/10.4995/riai.2024.19953}{https://doi.org/10.4995/riai.2024.19953}.}}
}

\maketitle

\begin{abstract}
Microfluidics, the study of fluids in microscopic channels, has led to important advances in fields as diverse as microelectronics, biotechnology and chemistry. Microfluidic research is primarily based on the use of microfluidic chips, low-cost devices that can be used to perform laboratory experiments using small amounts of fluid. These systems, however, require advanced control mechanisms in order to accurately achieve the flow rates and pressures required in the experiments. In this paper, we present the design of a model predictive controller intended to regulate the fluid flows in one of these systems. The results obtained, both through simulations and real experiments performed on the device, show that predictive control is an ideal technique to control these systems, especially taking into account all the existing constraints.
\end{abstract}


\section{Introduction}

The study of fluids under conditions where they flow through a network of micrometer-sized channels is known as microfluidics. Microfluidic systems are primarily based on the use of small chips, known as labs-on-a-chip, in which a variety of physical, chemical, or biological processes can be replicated. These systems have been applied in numerous fields such as chemistry, biotechnology, or medical research \citep{Ohno_08, gomezcontrolANESTESIA, ANESTESIA2MENDEZPEREZ2011241}. Microfluidics has made it possible, for example, to make advances in disease diagnosis or the simulation of organ behavior in small devices \citep{MF_ORGANS_CV19, bios12060370}. It also makes it possible to more realistically recreate the environment of certain microorganisms, allowing the study of more complex cultures. Another application where these devices stand out is in three-dimensional cell culture, where physiological conditions can be more reliably replicated \citep{duinen_microfluidic_2015}. Ultimately, this technology enables tests that provide more accurate results and reduce the need for animal experimentation. In addition, the use of these devices lowers the cost of carrying out certain experiments because the amount of chemicals required is very small, which also reduces the amount of waste generated. 

The dynamics of microfluidic systems typically depend on several variables. Therefore, to fully exploit their potential, it is essential to develop and implement advanced control methods. Specifically, model predictive control (MPC) is a potential control strategy for these devices, as it allows the control of systems with multiple inputs and multiple outputs. Its main advantage is that it allows predicting the evolution of the process in the future, taking into account all the physical constraints of the process in the calculation of the control action, whether they are in the input, state, output, or all of them simultaneously. However, the use of this type of control involves a more complex development and implementation than classical controllers, as it is essential to perform tasks such as building a model or defining the cost function to minimize. In the literature, very diverse applications of this type of control can be found, from its use in resonant power converters \citep{Resonantes}, to the cement industry \citep{1874indCementera}, improving the energy efficiency of air conditioning systems \citep{Marchante_Acosta_González_Zamarreño_Álvarez_2021}, controlling solar plants \citep{RIAISOLAR}, or clinical anesthesia applications \citep{gomezcontrolANESTESIA, ANESTESIA2MENDEZPEREZ2011241}, just to name a few examples. Additionally, more innovative developments are also underway, such as the combination of predictive control with neural networks for nonlinear systems \citep{calle_chojeda_control_2022}. All these references show that model predictive control is increasingly being used in numerous scientific and industrial fields.

Returning to the specific field of microfluidics, various works related to the development of controllers for these systems can be found in the literature. For example, in \cite{kim_modeling_2013}, a control system for a mechanical component of a microfluidic device is proposed, and in \cite{KuczenskiB617065J}, a procedure is described to control the interface between two fluids in a microscopic channel. However, most of these systems rely on the use of PID controllers, whose limitations in managing constraints are widely known. For example, \cite{heo_tuning-free_2016}, presents the design of a flow control algorithm and its comparison with a conventional PID controller, and \cite{B925497H} describes a PID controller for a microfluidic droplet generation device. The alternative proposed by some authors is in fact the development of model predictive control systems. Examples can be found in droplet-based microfluidic systems \citep{MADDALA2013132} or electroporation systems \citep{GhadamiComparativeStudy}, although none of these proposals specifically apply MPC techniques for flow control. Special mention should also be made of certain works, such as that of \citet{ArticuloDelITA}, focused specifically on the design of estimation algorithms intended for use in MPC controllers.

In this paper, we present the design of a model predictive controller for a microfluidic system. The objective is to control independently the flows circulating through three microscopic channels in a microfluidic chip, which meet at the same point to form a common outflow. This particular configuration is used in techniques such as \textit{Flow Focusing}, which consists in the generation of droplets and is used in fields such as drug encapsulation or particle design \citep{FF1,FF2}. The system is subject to numerous constraints, both on flow rates and applied pressures. Additionally, the coupling between different variables and the complexity of the system itself make it difficult to control using classical techniques, which justifies the need for MPC control. Since only partial measurements are available, the controller relies on a state observer whose design is also presented in the paper. The results obtained, both through simulation and on the real system, demonstrate the benefits of the proposed control system.

The structure of the article is summarized as follows. Following this introductory section, Section~\ref{sec:descripcion} describes the microfluidic system under study. Subsequently, in Section~\ref{sec:modelo}, we present the dynamic model of the complete system, as well as a simplified version used by the estimation and control algorithms. Sections~\ref{sec:observador} and~\ref{sec:mpc} describe, respectively, the state observer and the MPC controller designed. The results obtained with the complete implementation, first in simulation and then on the real system, are included in Section~\ref{sec:resultados}. Finally, Section~\ref{sec:conclusiones} presents the main conclusions of the work.

\section{System description}
\label{sec:descripcion}

\begin{figure}[t]
\centering
  \includegraphics[width=\linewidth]{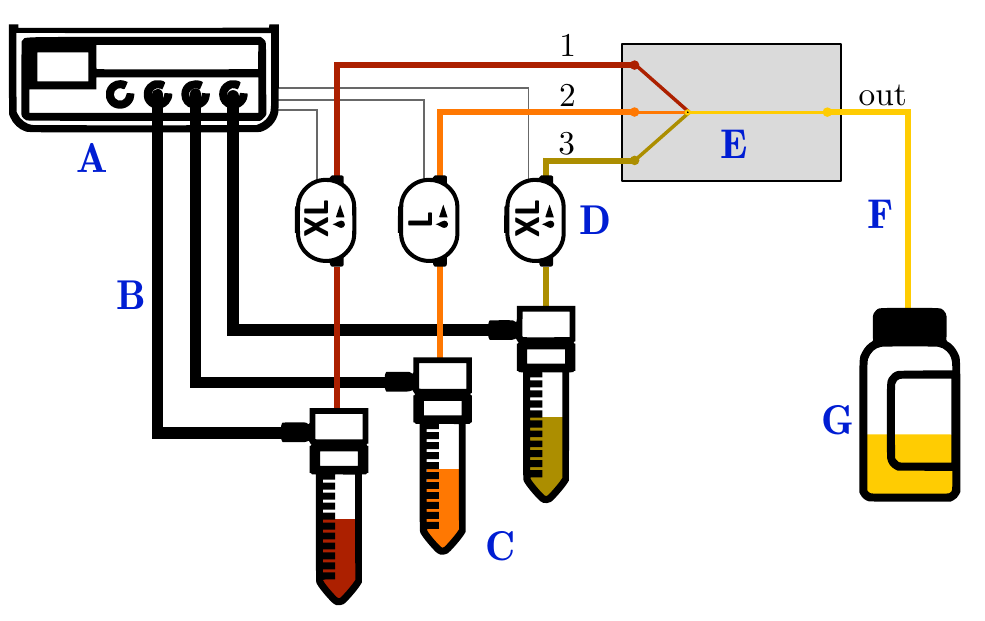}%
  \caption{Schematic diagram of the microfluidic system used.}
  \label{fig1}
\end{figure}

\begin{figure}[t]
\centering
  \includegraphics[width=6cm]{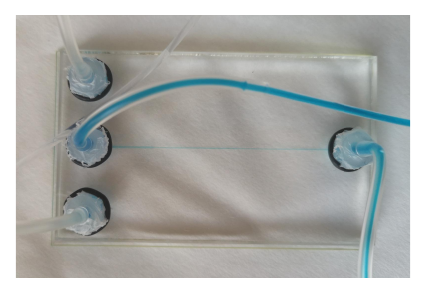}
  \caption{Photograph of the microfluidic chip.}\label{fig2}
\end{figure}

The microfluidic system used (see Fig.~\ref{fig1}) consists of three pressure regulators (A), three compressed air lines (B), three liquid reservoirs (C), three fluid lines with their respective flow meters (D), the microfluidic chip (E), an outlet line (F), and a final liquid reservoir (G). Fig.~\ref{fig2} shows a real image of the chip used.

\FloatBarrier

\begin{figure*}
\centering
    \includegraphics[width=17cm]{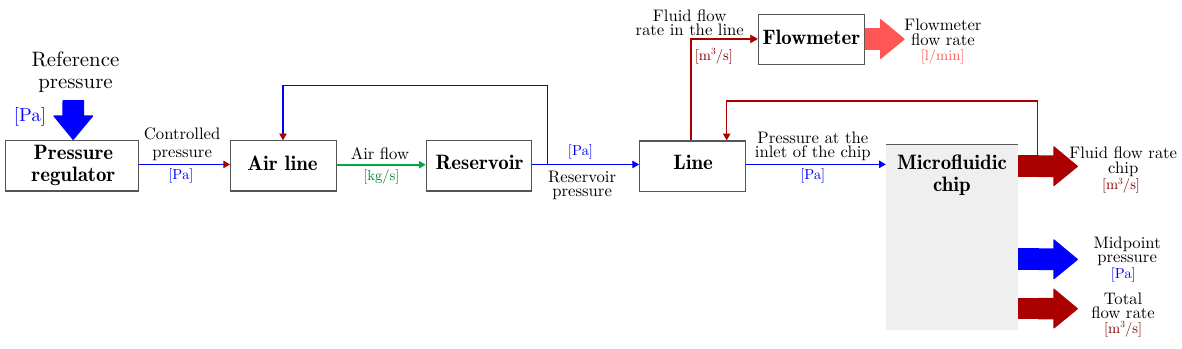}
    \caption{Conceptual block diagram for one of the lines.}
    \label{figdiagrama}
\end{figure*}

Fig.~\ref{figdiagrama} shows a conceptual block diagram for one of the lines, which is analogous to the three existing in the system. The system operation is as follows. The inputs are pressure references introduced to the regulators. These regulators act on the system to achieve, after a certain transient time, the requested pressures at the beginning of the compressed air lines. This pressure is transmitted to the reservoirs, which contain a fluid consisting of a mixture of water (74\% by volume) and glycerin (26\%), identical for all three lines. As a result of the pressure applied, the liquid contained in the reservoirs begins to flow through the fluid lines towards the chip. Next, the liquid enters the chip, where the three channels converge at a common point. Finally, all the fluid exits through the chip's outlet channel to a line that connects it to a reservoir at atmospheric pressure.  It is important to note that, in each of the three fluid lines between the reservoirs and the chip, the flow meters provide real-time measurements of the liquid flow rate. These measurements are the only ones used to close the control loop.

\section{System modeling}
\label{sec:modelo}

In this section, the equations of the dynamic model of the complete system presented in the previous section are presented. This model consists of the union of different submodels described in the following sections, each associated to a specific part of the system. It should be noted that a simplified version of this model is used for programming both the observer and the controller. The simplified model is described at the end of the section.

\subsection{Theoretical preliminaries}

The most relevant parameters in a microfluidic line are its hydraulic resistance, $\Res$, and inertia, $I$. The following expression can be used to calculate the inertia of a specific line,
\begin{equation}
    \label{eqn:Inertance}
    I = \frac{\rho \;l} {A} \;, 
\end{equation}
where $l$ and $A$ are, respectively, the length and cross-sectional area of the line, and $\rho$ is the fluid density. The hydraulic resistance of rectangular channels can be calculated using the following expression \citep{rapp_chapter_2017},
\begin{equation}
    \label{eqn:resistencia para un canal rectangular}
    \Res= \frac{12\,\mu\,l}{h^3\,w} \cdot \frac{1}{\left(1-\frac{192\,h}{\pi^5\,w}\right)\,\sum_{j=0}^\infty \left(\frac{1}{(2j+1)^5}\tanh\left(\frac{(2j+1)\,\pi\,w}{2\,h}\right)\right)},
\end{equation}
where $\mu$ is the dynamic viscosity of the fluid, $h$ is the height of the channel's cross-section, and $w$ is its width. On the other hand, for circular cross-section lines, the following expression is used \citep{rapp_chapter_2017},
\begin{equation}
    \label{eqn:resistencia para un canal circular}
    \Res = \mu \cdot \frac{8\,l}{\pi\,r^4},
\end{equation}
where $r$ is the radius of the cross-sectional area of the line.
 
Using Poiseuille's law and the hydraulic-electrical analogy, the flow variation in the channels can be formulated as a function of the pressure difference between two points in the channel, $\Delta P$, the hydraulic resistance, $\Res$, the flow rate through the channel, $Q$, and the fluid inertia in the channel, $I$. The hydraulic-electrical analogy allows some equations related to fluids to be reformulated by analogy with electrical equations. In this regard, an equivalence can be established between hydraulic resistance, $\Res$, and electrical resistance, $R_{e}$; between flow rate, $Q$, and electrical current, $I_{e}$; between the pressure difference between two points, $\Delta P$, and voltage drop, $V_{e}$; and between the fluid inertia in the channel, $I$, and the inductance, $L$, of an electrical circuit.
\begin{equation}
        \label{eqn:deltaPinicial}
    \Delta P = \Res\,Q + I\,\dot{Q}  \qquad \longleftrightarrow \qquad V_e = R_{e}\,I_e + L\,\dot{I}_e  \;
\end{equation}

Another relevant equation in these systems is the one describing the dynamics of pressures. The variation of pressure in a volume is expressed as the flow balance in that volume divided by the compressibility \citep{bruus2007theoretical},
\begin{equation}
     \label{eqn:def pm punto1}
    \dot{P}=\frac{\Delta Q}{C},
\end{equation}
where $\Delta Q$ is the difference between the flow entering and leaving this volume, and $C$ is the compressibility of the fluid in the chamber under study, given by
\begin{equation}
    \label{eqn:calculo de compresibilidad1}
    C = \frac{V_\tot}{E} , 
\end{equation}
where $V_\tot$ is the total volume of the chamber containing the fluid and $E$ is the bulk modulus of the fluid.

\subsection{Modeling of the microfluidic chip}

\begin{figure}[t]
\centering
   \includegraphics[width=7cm]{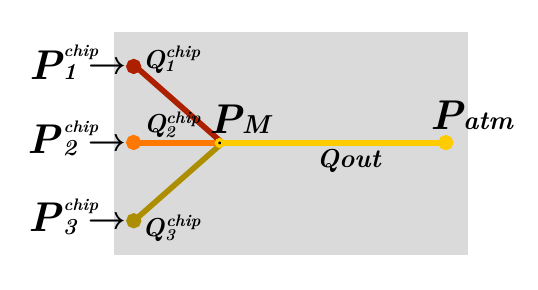}\\
  \caption{Chip diagram}\label{fig3}
\end{figure}

The chip (see Fig.~\ref{fig3}) consists of three rectangular cross-sectional input channels with their respective flow rates, $Q_1^\chip$, $Q_2^\chip$, and $Q_3^\chip$, and one output channel with flow rate $Q_\out$. The flow rates of the input lines are controlled by acting on the input pressures to the chip, $P_1^\chip$, $P_2^\chip$, and $P_3^\chip$. A variable $P_{M}$ is defined for the pressure at the midpoint where the three input channels merge. It should be noted that this variable cannot be measured directly. The output of the chip is at atmospheric pressure, $P_\atm$.

Applying \eqref{eqn:deltaPinicial} to each of the three input channels of the chip, the expressions of the derivatives of the flow rates can be obtained,
\begin{equation}
    \dot{Q}_{i}^\chip=\frac{P_{i}^\chip-P_{M}-\Res_{i}^\chip\,Q_{i}^\chip}{I_{i}^\chip}, \hspace{5mm} i \in \{1,2,3\}
    \label{eq:q_i_chip}
\end{equation}
where ${I_{i}^\chip}$ and ${\Res_{i}^\chip}$ are the inertia and hydraulic resistance of the $i$-th channel of the chip, calculated respectively using \eqref{eqn:Inertance} and \eqref{eqn:resistencia para un canal rectangular}. The dynamic equation for the output channel can be obtained similarly, in this case with atmospheric pressure at the outlet,
\begin{equation}
    \dot{Q}_\out=\frac{P_{M}-P_\atm-\Res_\out\;Q_\out}{I_\out},
\end{equation}
where $I_\out$ and $\Res_\out$ are respectively the inertia and hydraulic resistance of the output channel.

The compressibility of the fluid in the chip, $C^\chip$, can be obtained using \eqref{eqn:calculo de compresibilidad1}. In this case, the total volume is the sum of the volumes of each of the input channels, $V_i^{\chip}$, $i \in {1,2,3}$, and the output channel, $V_\out$.
\begin{equation}
    \label{eqn:calculo de compresibilidad}
    C^\chip = \frac{V_{1}^\chip + V_{2}^\chip + V_{3}^\chip + V_\out}{E}
\end{equation}
Besides, the pressure variation at the junction point of the lines can be obtained by applying \eqref{eqn:def pm punto1} to the chip.
\begin{equation}
     \label{eqn:def pm punto}
    \dot{P}_{M}=\frac{Q_{1}^\chip+Q_{2}^\chip+Q_{3}^\chip-Q_\out}{C^\chip}
\end{equation}

\subsection{Modeling of the fluid lines and flow meters}
\label{subsec:modelo_lineas_caudalim}

For the lines connecting the reservoirs to the chip, the flow meter is considered to be part of the line. This implies that, for each line, the inertia, volume, and resistance are the sum of those corresponding to the line and the flow meter.

The flow through the line depends on the pressure difference between the two ends of the line. From \eqref{eqn:deltaPinicial}, we obtain the following expressions for each of the three lines,
\begin{equation}
    \dot{Q}_{i}^{\lin}=\frac{P_{i}^{\,\res}-P_{i}^\chip-\Res_{i}^{\lin}\,Q_{i}^{\lin}}{I_{i}^{\lin}}, \ i \in \{1,2,3\},
\end{equation}
where $Q_{i}^{\lin}$ is the flow rate of the $i$-th fluid line, ${I_{i}^\lin}$ and ${\Res_{i}^\lin}$ are the inertia and hydraulic resistance of said line, calculated in this case using \eqref{eqn:Inertance} and \eqref{eqn:resistencia para un canal circular} for having a circular section, and $P_{i}^{\,\res}$ denotes the pressure of the $i$-th reservoir. The pressure variation at the end of each line can be obtained as a function of the flow rates by applying \eqref{eqn:def pm punto1} to this case,
\begin{equation}
     \dot{P}_{i}^\chip=\frac{Q_{i}^{\lin}-Q_{i}^\chip}{C_i^{\lin}}, \hspace{0.5cm} i \in \{1,2,3\},
\end{equation}
where $C_i^{\lin}$ is the compressibility of the fluid in the $i$-th line, obtained from \eqref{eqn:calculo de compresibilidad1} as
\begin{equation}
     C_i^{\lin} = \frac{V_i^{\lin} + V_i^{\fm} + V_i^{\chip}}{E} \approx \frac{V_i^{\lin} + V_i^{\fm}}{E}.
\end{equation}
In the previous expression, $V_i^{\lin}$ and $V_i^{\fm}$ are, respectively, the volumes of the $i$-th line and its corresponding flow meter. To calculate the fluid compressibility $C_i^{\lin}$, the influence of the volume of each of the chip channels, $V_i^{\chip}$, can be neglected as these are much smaller than those of the fluid lines and flow meters.

\subsection{Modeling of the reservoirs and air lines}

The three reservoirs containing the liquid are supplied with pressurized air through a series of tubes. This air causes the liquid to exit through the previously described lines. The ideal gas law is used to model the dynamics of this part,
\begin{equation}
    \label{eqn:eq estado gases ideales}
    \dot P_i^{\,\res} \, V_i^{\,\res} = \dot m_i \, R\,T,
    \vspace{3mm}
\end{equation}
where $\dot m_i$ is the mass flow rate of gas flowing through the $i$-th air duct and reaching each of the reservoirs, and $V_i^{\,\res}$ is the volume of the corresponding reservoir. Both the volume occupied by the air in each reservoir, $V_i^{\,\res}$, and the temperature, $T$, are assumed to be constant. For air, $R = 287.14$ J/(kg·K) is assumed.

The mass flow rate of air through the tube is calculated using the isentropic flow model with adiabatic index $\gamma=1.4$ \citep{cengel_mecanica_2018,white_mecanica_2008},
\begin{equation}\label{eqn:flujo aire}
    \dot{m}_i = \frac{A^{\duct}\,P_i^\reg}{\sqrt{R\,T}}\;\sqrt{7 \left( \frac{P_i^{\,\res}}{P_i^\reg} \right) ^{{10}/{7}} \left[1-\left( \frac{P_i^{\,\res}}{P_i^\reg} \right) ^{{2}/{7}} \right ]},
\end{equation}
where $A^{\duct}$ is the cross-sectional area of the duct, $P_i^\reg$ is the pressure at the beginning of the $i$-th air line (regulated by the corresponding pressure controller), and $P_i^{\,\res}$ is the pressure at the end of the line (pressure in the $i$-th reservoir).

\subsection{Modeling of the pressure regulators}

In the system used, there are three pressure regulators whose internal dynamics are a priori unknown. These regulators do not immediately provide the required pressure, but have their own dynamics that need to be modeled. In this case, we propose to use linear dynamic black-box models where the input is a pressure reference, ${u}_{i}$, and the output is the controlled pressure, $P_i^\reg$.

In order to identify these systems, the Matlab \textit{System Identification Toolbox} has been used to obtain a transfer function for each type of regulator by analyzing different sets of experimental data. The tests have been based on requesting different setpoints in the range of 0 to 150\,000~Pa and recording the step response of the regulators. By observing the system output, it was determined that the most appropriate model for the regulator of line~2 is a second-order linear model. For the other two lines, a different behavior was observed, which fits better with a third-order dynamics. That is, their dynamics are given by
\begin{align}
    \label{eqn:modeladoREGpresion1}
    {u}_{1} &= {P}_1^\reg + a_1\dot{P}_1^\reg + a_2\ddot{P}_1^\reg + a_3\dddot{P}_1^\reg,\\
    \label{eqn:modeladoREGpresion2}
    {u}_{2} &= {P}_2^\reg + b_1\dot{P}_2^\reg + b_2\ddot{P}_2^\reg,  \\
    \label{eqn:modeladoREGpresion3}
    {u}_{3} &= {P}_3^\reg + c_1\dot{P}_3^\reg + c_2\ddot{P}_3^\reg + c_3\dddot{P}_3^\reg,
\end{align}
where the coefficients $a_1$, $a_2$, $a_3$, $b_1$, $b_2$, $c_1$, $c_2$, and $c_3$ have also been estimated using the mentioned tool.

\subsection{Complete system model for simulation}

From all the previous equations, it is possible to build a dynamic model of the whole system with the 22 state variables listed in Table \ref{ve_todas}. This model, whose dynamics is nonlinear due to the use of equation \eqref{eqn:flujo aire}, has as inputs the three pressure references provided to the pressure regulators, and as outputs, the three flow readings from the three flow meters installed in the lines.

In order to perform realistic simulations, this model was implemented in Matlab-Simulink, taking into account certain effects such as possible noise in the measurements or the internal dynamics of the flow meters. This implementation has been used to validate through simulation both the observer and the predictive controller proposed in this work.

\FloatBarrier
\renewcommand{\arraystretch}{1.3} 
\begin{table}[b]
    \caption{State variables of the complete system.}
    \label{ve_todas}
    \begin{tabular*}{\hsize}{ll}
        \hline
        Name & Description  \\
        \hline \\[-1em]
        $Q_{1}^\chip$, $Q_{2}^\chip$, $Q_{3}^\chip$   & Flow rates at the chip inlet \\[0.6ex]
        $Q_\out$ & Flow rate at the chip outlet \\[0.6ex]
        $P_{M}$ & Pressure at the midpoint \\[0.6ex]
        $Q_{1}^{\lin}$, $Q_{2}^{\lin}$, $Q_{3}^{\lin}$ & Flow rates in the lines \\[0.6ex]
        $P_{1}^\chip$, $P_{2}^\chip$, $P_{3}^\chip$  & Pressures at the chip inlet \\[0.6ex]
        ${P}_1^{\,\res} $, ${P}_2^{\,\res} $, ${P}_3^{\,\res} $ & Pressures in the reservoirs \\[0.6ex]
        ${P}_1^\reg,\dot{P}_1^\reg,\ddot{P}_1^\reg $ & Controlled pressure 1 and its derivatives \\[0.6ex]
        ${P}_2^\reg,\dot{P}_2^\reg $ & Controlled pressure 2 and its derivative \\[0.6ex]
        ${P}_3^\reg,\dot{P}_3^\reg,\ddot{P}_3^\reg $ & Controlled pressure 3 and its derivatives \\[0.6ex]
        \hline
  \end{tabular*}
\end{table}
\renewcommand{\arraystretch}{1}

\subsection{Simplification for observation and control}
\label{subsec:modelo_simple}

Based on the complete model, a simplification has been performed by grouping the whole system into a lower-order state-space model with linear dynamics. In this regard, it has been verified by simulation that neglecting the effect of the air reservoir and possible losses in it has minimal influence. Specifically, for a set point of 10\,000~Pa applied equally to all three lines, the difference between the steady-state flow rates is less than $0.5 \%$, calculated as the difference between the flow rate of the two models with respect to the flow rate of the complete system. This allows to reduce the complexity of the model used in the design of the observer and the controller by simplifying the nonlinear part, while leaving the rest with linear dynamics. Thus, the pressures ${P}_i^\reg$ and ${P}_i^{\,\res}$ are assumed to be equal for each of the three lines. It is also assumed that the flow rate through each line is equal to the flow rate through the corresponding channel inside the chip, i.e., $Q_{i}^{\lin} = Q_{i}^\chip$. Since the flow meter is considered to be part of the line (see section \ref{subsec:modelo_lineas_caudalim}), the flow rate measured by it will directly correspond to the flow through the corresponding line and inside the chip.

Considering the above, the lines linking the reservoirs and the chip and the chip channels themselves are considered in the simplified model as a set, thus simplifying the pressures at the chip inlet, $P_{i}^\chip$. That is, the simplified model assumes that the pressure at the inlet of the system composed of one of the lines and the corresponding internal channel of the chip is the controlled pressure ${P}_i^\reg$. In making this union, the resistance and inertia properties of the assembly are given by
\begin{equation}
\label{eqn:parameteros_juntos_simplificado}
\left\{
\begin{array}{r@{}l}
\Res_i \ & =(\Res_{i}^\chip+\Res_{i}^{\lin})   \\
I_i \ & =(I_{i}^\chip+I_{i}^{\lin})   
\end{array} \right. \;\;\;\;\;\; i  \in \{1,2,3\}.
\end{equation}
With the simplification, the variables for intermediate pressures and flow rates in the lines are no longer used, reducing the number of state variables from 22 to 13 (see Table \ref{ve_simplificado}). Furthermore, in subsequent analyses, to simplify the operations, all pressures are considered relative to atmospheric pressure, so $P_\atm = 0$~Pa is assumed in the model. The resulting model can be represented compactly as
\begin{align}
    \mathbf{\dot x_m}(t) &= \mathbf{A_m}\,\mathbf{x_m}(t) + \mathbf{B_m}\, \mathbf{u}(t), \\
    \mathbf{y}(t) &= \mathbf{H_m}\,\mathbf{x_m}(t),
\end{align}
where $\mathbf{x_m} \in \mathbb{R}^n$, with $n=13$, is the state vector, formed by the state variables from Table~\ref{ve_simplificado} (in that order), and $\mathbf{u}\in\mathbb{R}^m$, with $m=3$, and $\mathbf{y}\in\mathbb{R}^p$, with $p=3$, are, respectively, the input and output vectors of the system, defined as:
\begin{equation*}
    \mathbf{u} = \left[u_1\;\ u_2\;\ u_3\right]^T, \hspace{1em} 
    \mathbf{y} = \left[y_1\;\ y_2\;\ y_3\right]^T = \left[Q_1\;\ Q_2\;\ Q_3\right]^T.
\end{equation*}
The state ($\mathbf{A_m}$), input ($\mathbf{B_m}$), and output ($\mathbf{H_m}$) matrices of the model are not explicitly shown in this article due to space limitations. However, they can be directly obtained from the equations \eqref{eq:q_i_chip}--\eqref{eqn:parameteros_juntos_simplificado}.

Regarding the accuracy of the simplification made, it has been verified by simulation that, given a constant set point of 150\,000~Pa (the maximum supported by the system), the maximum difference in steady state between the flow rates of the complete and simplified models is less than 6.5\% with respect to the system without simplification. During the transient response, there are slightly larger differences, but we have considered that the behavior is acceptable enough to use the simplified model.

This simplified model is used in the design of the controller and observer. Specifically, the implementation uses the discrete-time version of the model,
\begin{align}
    \mathbf{x_m}(k+1) &= \mathbf{F_m}\,\mathbf{x_m}(k) + \mathbf{G_m}\, \mathbf{u}(k), \label{eq:xm_discreto}\\
    \mathbf{y}(k) &= \mathbf{H_m}\,\mathbf{x_m}(k), \label{eq:ym_discreto}
\end{align}
where $\mathbf{F_m}$ and $\mathbf{G_m}$ are, respectively, the state transition matrix and the input matrix in discrete time.

\renewcommand{\arraystretch}{1.3} 
\begin{table}[t]
    \caption{State variables of the simplified system.}
    \label{ve_simplificado}
    \begin{tabular*}{\hsize}{ll}
        \hline
        Name & Description  \\
        \hline \\[-1em]
        $Q_{1}$, $Q_{2}$, $Q_{3}$ & Flow rates in the line + channel assembly \\[0.6ex]
        $Q_\out$ & Flow rate at the chip outlet \\[0.6ex]
        $P_{M}$ & Pressure at the midpoint \\[0.6ex]
        ${P}_1^\reg,\dot{P}_1^\reg,\ddot{P}_1^\reg $ & Controlled pressure 1 and its derivatives \\[0.6ex]
        ${P}_2^\reg,\dot{P}_2^\reg $ & Controlled pressure 2 and its derivatives \\[0.6ex]
        ${P}_3^\reg,\dot{P}_3^\reg,\ddot{P}_3^\reg $ & Controlled pressure 3 and its derivatives \\[0.6ex]
        \hline
    \end{tabular*}
\end{table}
\renewcommand{\arraystretch}{1}

\section{State observer}
\label{sec:observador}

\begin{figure*}
\centering
    \includegraphics[width=1\linewidth]{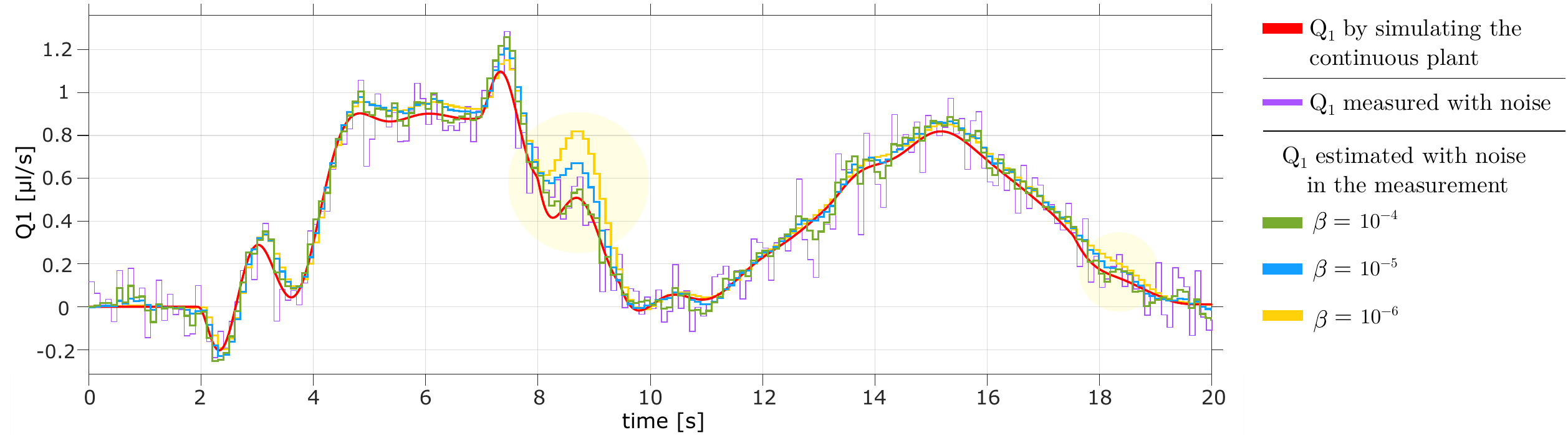}%
    \caption{Simulation results of the Kalman filter performance in the presence of noise.}
    \label{graficaKalman}
\end{figure*}

\subsection{Observer design}

The simplified model described previously has thirteen state variables, whose values need to be known in real time in order to control the system. In practice, however, it is not possible to measure all of them since the three available flow meters only allow for the measurement of three of these state variables. It is thus necessary to design and implement an observer that estimates the unmeasured variables in order to reconstruct the full state at each instant. In this work, the state observation is carried out using a Kalman filter. The filter is based on the classical formulation in \cite{kalman_new_1960} and \cite{welch_introduction_2006}, and it is described by the following equations:
\begin{align}
    {\mathbf{x}'_\mathbf{m}}(k) &=  \mathbf{F_m} \, \hat{\mathbf{x}}_\mathbf{m}(k-1) + \mathbf{G_m} \, \mathbf{u}(k-1),\\
    {\mathbf{P}'}(k) &=  \mathbf{F_m} \, \hat{\mathbf{P}}(k-1) \, \mathbf{F_m}^{T} + \Qkf,\\
    {\mathbf{K}}(k) &=  \mathbf{P'}(k) \, \mathbf{H_m}^T \left[\mathbf{H_m} \, \mathbf{P'}(k) \,  \mathbf{H_m}^T +  \Rkf\right]^{-1},\\
    {\hat{\mathbf{x}}}_\mathbf{m}(k) &=  \mathbf{x}'_\mathbf{m}(k) + \mathbf{K}(k) \left[\mathbf{y}(k) - \mathbf{H_m} \, \mathbf{x}'_\mathbf{m}(k)\right], \\
    {\hat{\mathbf{P}}}(k) &=  \mathbf{P'}(k) - \mathbf{K}(k) \, \mathbf{H_m} \, \mathbf{P'}(k),
\end{align}
where ${\mathbf{x}'_\mathbf{m}}$ and $\hat{\mathbf{x}}_\mathbf{m}$ are the \textit{a priori} and \textit{a posteriori} estimates of the state ${\mathbf{x}_\mathbf{m}}$, ${\mathbf{P}'}$ and $\hat{\mathbf{P}}$ are the corresponding covariance matrices of these estimates, and $\mathbf{K}$ is the Kalman gain. Additionally, there are parameters that are crucial for the filter's performance: the covariance matrices of the measurement noise, denoted as $\Rkf\in \mathbb{R}^{p \times p}$, and the process noise, denoted as $\Qkf\in \mathbb{R}^{n \times n}$.

\subsection{Simulation and tuning of the observer}

In order to verify the correct design of the filter, several simulations have been carried out in the software Matlab-Simulink, with the complete model playing the role of the real system, modeled as previously indicated. However, the observer is based on the simplified model. Thus, in the simulation, the model of the system available to the observer is not exactly the same as that of the plant.

Based on preliminary tests on the system, it was determined that the measurement noise of the flow meters is in the range of $\pm \; 0.3$ $\mu$l/s. Assuming Gaussianity and that this value corresponds to three standard deviations, it was concluded that the variance of the measurement noise should be about $10^{-20}$~(m$^3$/s)$^2$. Consequently, considering that the three sensors are independent and thus the measurements must be uncorrelated, we decided to set $\Rkf = \mathbf{I}_p \cdot 10^{-20}$, where $\mathbf{I}_p$ is the identity matrix of size~$p=3$.

For the definition of the matrix $\Qkf$ it is assumed that the noise affecting each state variable in \eqref{eq:xm_discreto} is independent of the others. In other words, $\Qkf$ is set to be diagonal. Furthermore, to determine its value, it has been assumed that such noise signals are Gaussian with zero mean and standard deviation proportional to a characteristic value of the corresponding variable. Therefore, the covariance of the process noise has the form
\begin{align}
    \Qkf &= 
    \begin{bmatrix} 
        \Qkf_Q & \mathbf{0}_{4\times 9} \\ 
        \mathbf{0}_{9\times 4} & \Qkf_P
    \end{bmatrix},
\end{align}
where 
\begin{equation*}
    \Qkf_Q = \beta \cdot
    \begin{bmatrix} 
        1 & 0 & 0 & 0 \\ 
        0 & 1 & 0 & 0 \\ 
        0 & 0 & 1 & 0 \\ 
        0 & 0 & 0 & 9
    \end{bmatrix}\cdot10^{-18}
\end{equation*}
is the submatrix of $\Qkf$ associated with the flow rate variables, {$\Qkf_P = \beta \cdot \mathbf{I}_9 \cdot10^{8}$} is the submatrix corresponding to the pressures and their derivatives, $\beta$ is a dimensionless factor that allows adjusting the filter's behavior, and $\mathbf{0}_{p\times n}$ is a matrix of zeros with $p$ rows and $n$ columns. Note that the different orders of magnitude arise from working with different magnitudes (pressures and flow rates) in very different numerical ranges.

When setting the value of $\beta$, a decision must be made whether to minimize the effect of noise, relying solely on the model of the plant available to the filter and sacrificing some measurement information, or to allow a minimal level of noise effect and improve the response to discrepancies in the plant. Fig.~\ref{graficaKalman} shows the evolution of the flow rate $Q_1$ for different values of $\beta$. In all these simulations, measurement noise with the mentioned variance, i.e., $10^{-20}$~(m$^3$/s)$^2$, has been artificially introduced. Successive simulations have allowed us to determine that the value of $\beta=$ {$10^{-4}$} preserves the real behavior with a reduced noise effect. For higher values, too much confidence is placed in the simplified model, leading to errors (areas highlighted in yellow in the figure), while for lower values, too much noise is observed.

The simulations performed lead to the conclusion that the designed observer is valid for observing the system state variables even with some measurement noise. Based on the results, it is determined that in order to obtain the best performance, it would be advisable to adjust the matrices during the system calibration to best match the specific operating conditions. In cases where the system undergoes significant changes during operation, it would be appropriate to use an adaptive algorithm that modifies the model matrices according to the observations.

\section{MPC controller}
\label{sec:mpc}

\subsection{Controller design}

The following is a description of the model predictive controller (MPC) design used for this system. One of the main strengths of this type of control is that the formulation allows for controlling multi-input, multi-output systems, including constraints on the input, state, or output in the formulation. Additionally, MPC allows for real-time optimization, explicitly using a plant model to predict the process evolution in future time steps by minimizing an objective function.

The formulation used in the implementation of the MPC controller in this system is based on that described in \citet{wang_model_2009} and \citet{Camacho_Bordons_2010}. This formulation utilizes the following extended model of the system,
\begin{align}
    \mathbf{x}(k+1) &= \mathbf{F}\,\mathbf{x}(k) + \mathbf{G}\,\Delta\mathbf{u}(k), \label{eq:mod_ampl_x} \\
    \mathbf{y}(k) &= \mathbf{H}\,\mathbf{x}(k), \label{eq:mod_ampl_y}
\end{align}
where $\mathbf{x}$ is the extended state, defined as
\begin{equation}
    \mathbf{x}(k) =
    \begin{bmatrix} 
        \Delta \mathbf{x}_{\mathrm{m}}(k) \\ 
        \mathbf{y}(k) 
    \end{bmatrix} =
    \begin{bmatrix} 
        \mathbf{x}_{\mathrm{m}}(k)-\mathbf{x}_{\mathrm{m}}(k-1) \\ 
        \mathbf{y}(k) 
    \end{bmatrix}.
\end{equation}
It is important to note that the input to this model consists of control action increments, that is, $\Delta \mathbf{u}(k) = \mathbf{u}(k) - \mathbf{u}(k-1)$, rather than the control actions themselves. This implies that the MPC formulation implicitly includes an integrator for each output, allowing the desired setpoint to be reached in steady-state even in the presence of disturbances or modeling errors. The matrices $\mathbf{F}$, $\mathbf{G}$, and $\mathbf{H}$, which can be obtained by operating on \eqref{eq:xm_discreto}--\eqref{eq:ym_discreto}, have the following form:
\begin{equation*}
    \mathbf{F} = 
    \begin{bmatrix} 
        \mathbf{F}_{\mathrm{m}} & \mathbf{0}^{T}_{p\times n} \\ 
        \mathbf{H}_\mathrm{m} \mathbf{F}_\mathrm{m} & \mathbf{I}_{p}
    \end{bmatrix}, \hspace{0.3em}
    \mathbf{G} =
    \begin{bmatrix} 
        \mathbf{G}_{\mathrm{m}} \\ 
        \mathbf{H}_\mathrm{m}\mathbf{G}_\mathrm{m}
    \end{bmatrix}, \hspace{0.3em}
    \mathbf{H} =
    \begin{bmatrix} 
        \mathbf{0}_{p\times n} & \mathbf{I}_{p}
    \end{bmatrix}.
\end{equation*}
In this work, we consider the control horizon and the prediction horizon to be equal, which we will simply refer to as the horizon and denote as $N$. Based on this horizon, it is possible to define the following vectors:
\begin{equation}
    \label{eqn: Def deltaU }
    \Delta \mathbf{U}(k) = 
    \begin{bmatrix} 
        \Delta \mathbf{u}(k)\\[-3pt]
        \vdots \\ 
        \Delta \mathbf{u}(k+N-1)
    \end{bmatrix} 
\end{equation}

\begin{equation}
    \mathbf{Y}(k) = 
    \begin{bmatrix} 
        \mathbf{y}(k+1\,|\,k)\\[-3pt]
        \vdots \\ 
        \mathbf{y}(k+N\,|\,k)
    \end{bmatrix}
    \quad  \mathbf{Y_d}(k) = 
    \begin{bmatrix} 
        \mathbf{y_d}(k+1)\\[-3pt]
        \vdots \\ 
        \mathbf{y_d}(k+N)
    \end{bmatrix},
\end{equation}
where $\mathbf{y}(j\,|\,k)\in\mathbb{R}^p$ is the prediction of the system output at the $j$-th instant based on the information available at the $k$-th instant, and $\mathbf{y_d}(j)\in\mathbb{R}^p$ is the desired value for that output at that instant. In other words, the vectors $\Delta \mathbf{U} \in \mathbb{R}^{Nm}$, $\mathbf{Y} \in \mathbb{R}^{Np}$ and $\mathbf{Y_d} \in \mathbb{R}^{Np}$ contain, respectively, the control action increments, the predicted outputs, and the desired values for the whole horizon. Operating with these vectors and the model \eqref{eq:mod_ampl_x}--\eqref{eq:mod_ampl_y}, it is possible to express the vector $\mathbf{Y}(k)$ as
\begin{equation}
    \mathbf{Y}(k) = \mathbf{\Psi}\,\mathbf{x}(k) + \mathbf{\Phi} \,\Delta \mathbf{U}(k),
\end{equation}
where $\mathbf{\Psi}$ and $\mathbf{\Phi}$ are constant matrices defined as
\begin{equation*}
    \label{eqn: Def F y phi}
    \mathbf{\Psi} \! = \!\!
    \begin{bmatrix} 
        \mathbf{H}\mathbf{F}\\
        \mathbf{H}\mathbf{F}^2 \\[-3pt]
        \vdots \\[2pt] 
        \mathbf{H}\mathbf{F}^N
    \end{bmatrix}\!,\,
     \mathbf{\Phi} \! = \!\!
    \begin{bmatrix} 
        \mathbf{H}\mathbf{G} & \mathbf{0}_{p\times m} &  \cdots & \mathbf{0}_{p\times m} \\ 
        \mathbf{H}\mathbf{A}\mathbf{G} & \mathbf{H}\mathbf{G} &  \cdots & \mathbf{0}_{p\times m} \\[-3pt]
        \vdots & \ddots& \ddots &   \\[2pt] 
        \mathbf{H}\mathbf{F}^{N-1}\mathbf{G} & \mathbf{H}\mathbf{F}^{N-2}\mathbf{G}  & \cdots & \mathbf{H}\mathbf{G}
    \end{bmatrix}\!.
\end{equation*}

Based on the above definitions, the cost function to be minimized, $J$, is formulated as the sum of a tracking error term, which takes into account how far the flow rates are from the desired values, and another term associated with the increase in the control variable,\begin{equation}
    \label{eqn:funcion coste1mem}
    J(k) = \tilde{\mathbf{Y}}(k)^T \cdot \Qmpc \cdot \tilde{\mathbf{Y}}(k) + \Delta \mathbf{U}(k)^T \cdot \Rmpc \cdot \Delta \mathbf{U}(k) ,
\end{equation}
where $\tilde{\mathbf{Y}} = \mathbf{Y} - \mathbf{Y_d}$ is the vector with the tracking errors. The diagonal matrix $\Qmpc \in \mathbb{R}^{Np\, \times\, Np}$ determines the importance of the error $\tilde{\mathbf{Y}}$ in $J$. Additionally, it also establishes the relative weights between the errors of the three outputs. In this case, the three outputs are flow rates that are within the same value ranges. Since the importance of the error is the same for all three, we decided to set all these values to unity. That is, $\Qmpc=\mathbf{I}_{Np}$. On the other hand, the diagonal matrix $\Rmpc \in \mathbb{R}^{Nm \, \times \, Nm}$ allows to define the relative weights of the control actions, both among themselves and with respect to the term associated with the error $\tilde{\mathbf{Y}}$. Since in the system under studye the three actions are pressures within the same value ranges, and the relative importance is the same for all three, we decided that all the elements of the diagonal take the same value. That is, $\Rmpc=\alpha\,\mathbf{I}_{Nm}$, where $\alpha$ is a parameter that allows to adjust the behavior of the controller.

The controller formulation also includes constraints on the control actions (the pressures to be applied) both in their magnitude and in their rate of change, as well as on the three flow rates to control. The implementation requires that all constraints be expressed in terms of increments of the controlled variable, according to the following formulation,
\begin{equation}
    \label{eqn:FormulacionRestriccionesMPC}
    \mathbf{M}\cdot \mathbf{\Delta U}(k) \leq \mathbf{\gamma}(k),
\end{equation}
where
\begin{align}
    \mathbf{M}&=
    \begin{bmatrix}
    -\mathbf{C_2}\\
    \mathbf{C_2}\\
    -\mathbf{I}_{m  N}\\
    \mathbf{I}_{m  N}\\
    -\mathbf{\Phi}\\
    \mathbf{\Phi}\\
    \end{bmatrix},
    &
    \mathbf{\gamma}(k) &= 
    \begin{bmatrix}
    -\mathbf{U}_{min}+\mathbf{C_1}\, \mathbf{u}(k-1)\\
    +\mathbf{U}_{max}-\mathbf{C_1}\, \mathbf{u}(k-1)\\
    -\mathbf{\Delta U}_{min}\\
    +\mathbf{\Delta U}_{max}\\
    -\mathbf{Y}_{min}+\mathbf{\Psi}\, \mathbf{x}(k)\\
    +\mathbf{Y}_{max}-\mathbf{\Psi}\, \mathbf{x}(k)\\
    \end{bmatrix}. \nonumber
\end{align}
The matrix $\mathbf{M}$ establishes the relationship between the constraints and the increment of the control variable, and the vector $\mathbf{\gamma}$ formulates each of the constraints in terms of the control variation. The vectors $\mathbf{U}_{min}$, $\mathbf{U}_{max}$, $\mathbf{\Delta U}_{min}$, $\mathbf{\Delta U}_{max}$, $\mathbf{Y}_{min}$ and $\mathbf{Y}_{max}$ set the maximum and minimum bounds for the actions, the rate of change of the actions, and the outputs, respectively. On the other hand, $\mathbf{C_1}$ and $\mathbf{C_2}$ are auxiliary matrices formed by zeros and ones in the appropriate arrangement to relate the constraints to the increment of the control variable.

Once the cost function to be minimized and the constraints to be satisfied are defined, the problem becomes an optimization of a quadratic function with constraints.
\begin{align}
     \underset{\Delta \mathbf{U}(k)}{\mbox{min.}} \; J(k)     \hspace{5mm}
    \mbox{st:} \; \; \eqref{eqn:FormulacionRestriccionesMPC}
    \label{eqn: Def minim}
\end{align}%

\subsection{Controller simulation and tuning}

The goal of the simulations presented here is to verify the correct design of the controller. Thus, these have been carried out without using the observer, assuming that the full state is measurable. The operation of the complete system including the observer will be analyzed in the following section.

The simulations have been carried out using Matlab-Simulink. Specifically, the optimization problem \eqref{eqn: Def minim} has been solved using the Matlab optimization toolbox, with the \textit{quadprog} function designed for this purpose. When applying this function, the vector of control action increments, $\Delta \mathbf{U}$, to be applied for all subsequent time steps of the control horizon is obtained. However, according to the paradigm of model predictive control, only the calculated input for the next time step will be applied, redoing this calculation at each step.

The choice of the sampling frequency and the discretization of the controller has taken into account the direct connection between the sampling period, $T$, the number of steps in the horizon, $N$, and the duration of the horizon, $T\cdot N$. This implies a trade-off between computational cost and controller performance. For their selection, an iterative process was carried out, resulting in a horizon of $N=10$ steps and a sampling period of $T=0.1$ seconds.

\begin{figure}[p]
\centering
  \includegraphics[width=0.7\linewidth]{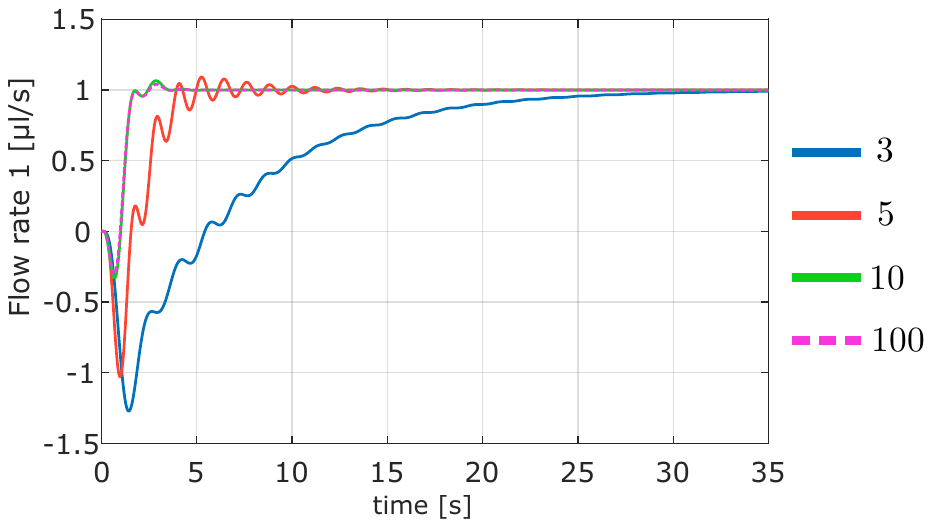}%
  \caption{Simulated response of the flow rate $Q_1$ for $T=0.1$~s and different values of the horizon $N$. The setpoint is 1 $\mu$l/s and $\alpha=10^{-7}$.}
  \label{fig:ComparacionHorizontes}
\vspace{0.8\floatsep}
  \includegraphics[width=0.95\linewidth]{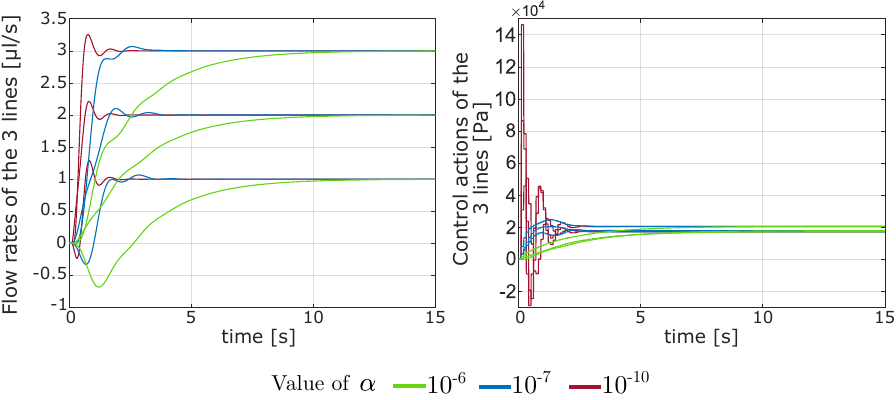}%
  \caption{Simulated system response for $T=0.1$~s, horizon $N=10$, and different values of the parameter $\alpha$.}
  \label{fig:RespuestaDiferentesR}
\vspace{0.8\floatsep}
  \includegraphics[width=0.95\linewidth]{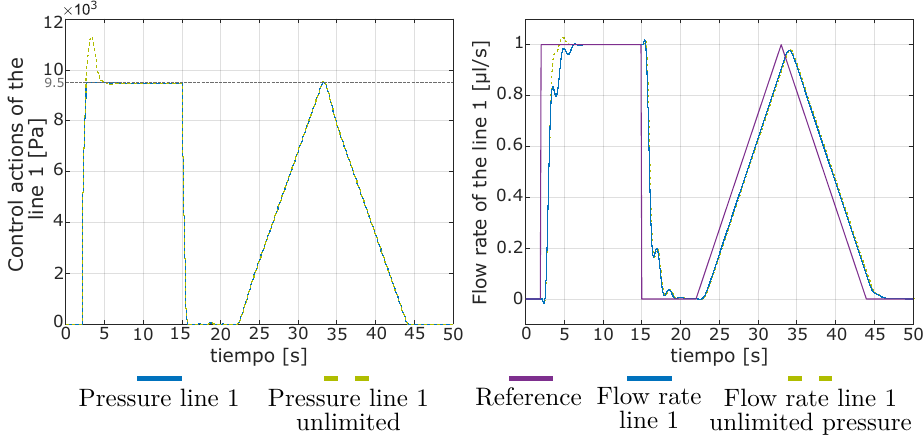}%
  \caption{Simulation under constraints on the inputs.}
  \label{fig:ResPresion}
\vspace{0.8\floatsep}
  \includegraphics[width=0.95\linewidth]{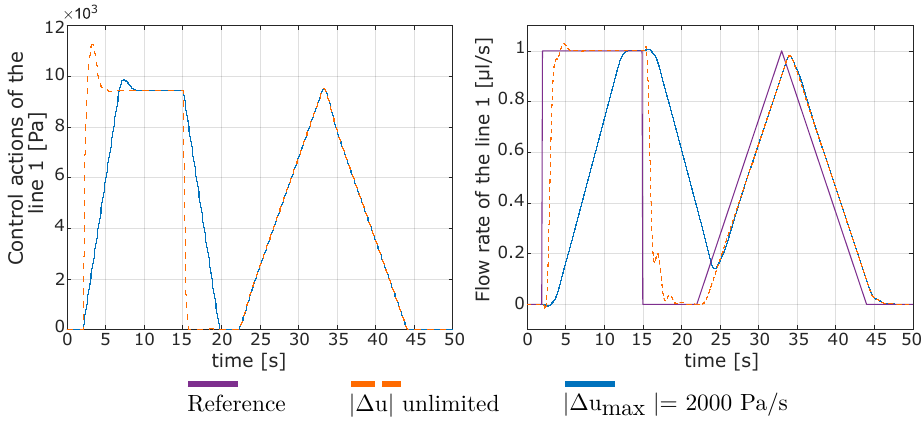}%
  \caption{Simulation under constraints on the variation of the inputs.}
  \label{fig:ResDelta}
\vspace{0.8\floatsep}
  \includegraphics[width=0.95\linewidth]{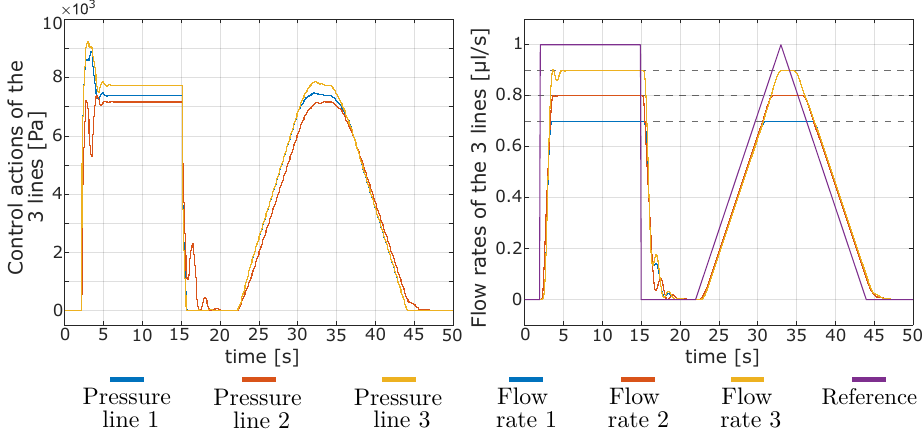}%
  \caption{Simulation under constraints on the flow rates.}
  \label{fig:ResCaudalesSalida}
\end{figure}

In order to demonstrate the validity of these values, several simulations were carried out without restrictions on any of the variables, setting a reference of 1 $\mu$l/s for flow~$Q_1$, 2 $\mu$l/s for flow~$Q_2$, and 3 $\mu$l/s for flow~$Q_3$. The result for different horizons can be observed in Fig.~\ref{fig:ComparacionHorizontes}. Only the result for line~1 is shown, as it is similar for all three. It is verified that the best value is $N=10$, since a further increase increase only provides a higher computational cost without any benefit in the behavior. This will be the horizon used in the remaining cases. With the horizon fixed, Fig.~\ref{fig:RespuestaDiferentesR} shows the system behavior for different values of the controller tuning parameter $\alpha$. It can be observed that $\alpha = 10^{-7}$ results in the best trade-off between response time and overshoot.

Additional simulations implementing different constraints have also been conducted. The performance of the control system has been verified for three different cases. In the first case (see Fig.~\ref{fig:ResPresion}), a limit of 9\,500~Pa has been set on the pressure to apply. The goal is to prevent a peak that appears in the control action when a step response is asked. By not allowing this peak, the response becomes slightly slower, but the flow rate still reaches the desired value. In the second case (see Fig.~\ref{fig:ResDelta}), the response to a limitation of the action variation to 2\,000~Pa/s has been analyzed. When a step input is applied, the response becomes slower but reaches the reference. With a ramp input, since the slope is not steep enough, the behavior is identical to the case without limitations. Finally, the performance has been examined under different constraints for each of the input flow rates to the chip, causing the controller to adjust the pressures to meet the constraints. This last case is shown in Fig.~\ref{fig:ResCaudalesSalida}. Based on these results, it is concluded that the controller design is correct.

\section{Results of the complete implementation}
\label{sec:resultados}

After presenting the model, the observer, and the controller separately, this section reports the results obtained from tests conducted using the complete control system, both through simulation and in experiments carried out on the actual microfluidic system.

\subsection{Validation by simulation}

\begin{figure*}
\centering
    \includegraphics[width=\linewidth]{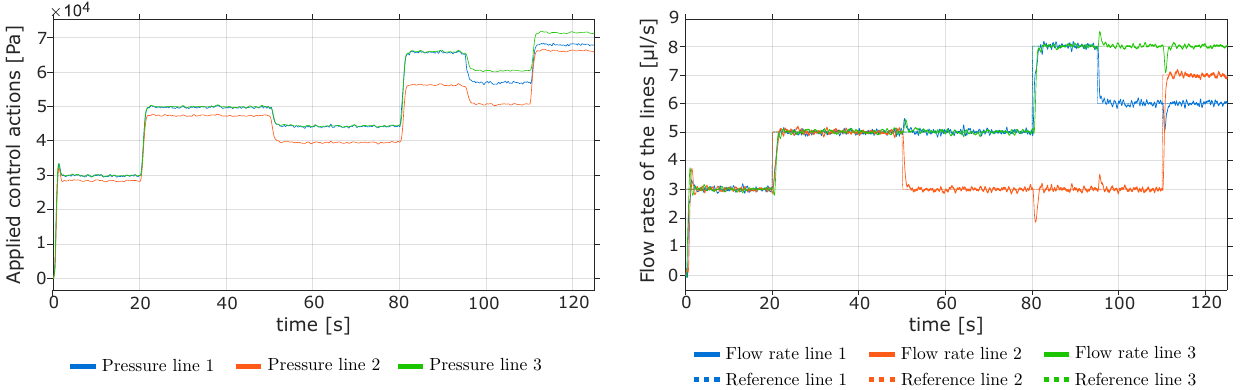}%
    \caption{Simulation of the complete MPC system. Pressures (left) and flow rates (right) in response to stepped setpoints, different for each line.}\label{fig:MPC+KFqsRUIDO}
\end{figure*}

Matlab-Simulink has been used to simulate the most realistic conditions. As described earlier, the plant has been modeled in continuous time and, accordingly, it is simulated using a precise numerical integration method that provides realistic results of the physical system response. Working in a simulation environment allows for measuring all state variables. It should be noted, however, that only the readings from the flow meters are used to provide feedback to the controller. These readings, to which simulated measurement noise can be added, serve as inputs to the observer, which estimates the rest of the state variables needed by the MPC algorithm. The controller is also provided with the desired values for the flow rates of the three lines. The controller performs the operations, and once the control action is calculated, it is applied to the plant, closing the loop. The simulations have been performed with a horizon $N=10$, a sampling period $T=0.1$~s, and values of $\alpha = 10^{-7}$ and $\beta = 10^{-4}$ for the controller and observer tuning parameters, respectively.

For the constraints to be imposed on the controller, the following considerations have been taken into account:
\begin{enumerate}
    \item The maximum pressure of the controllers is limited by the pressure that the glass containers used in the experiments can withstand (150\,000~Pa relative to atmospheric pressure). The minimum pressure is limited to atmospheric pressure.
    \item The maximum rate of change has been determined based on the data provided from previous experiments. It is set at 100\,000~Pa/s.
    \item The only constraint on the flow rate in the lines entering the chip is that it must be positive, to prevent fluid from one line from entering another.
\end{enumerate}

Fig.~\ref{fig:MPC+KFqsRUIDO} shows the flow rates and pressures of the lines for a simulation in which stepped setpoints have been established, different for each of the flow rates. As in the observer validation, noise with variance $10^{-20}$~(m$^3$/s)$^2$ has also been added to the three flow meters measurements. It can be seen that the system has a response time of approximately 1.8 seconds for lines~1 and~3, and 1.5 seconds for line~2. The results also show that the system correctly reaches the desired values with a small overshoot, which confirms that it works properly.

\subsection{Validation on the real system}

To conduct tests on the real system, the observer and controller code has been ported from Matlab to Python. The manufacturer of the devices used (Fluigent) provides a set of tools to easily communicate and control these devices from Python code, which is why this language has been chosen. In the following, some of the results obtained in the experiments are presented. The general constraints imposed are the same as in the simulations shown earlier, although for safety reasons the pressures on the real system have been kept below 100\,000~Pa.

The parameters of the MPC controller are the same as those of the previous simulations. That is, $N=10$, $T=0.1$~s, and $\alpha = 10^{-7}$. However, after some preliminary tests, it was observed that the real system exhibited more differences than expected from the model, especially in line~1, where an unexpected oscillatory dynamics was detected. Thus, to deal with this uncertainty, it was considered that the matrix $\Qkf$ of the filter required some manual adjustment. The values of each of the submatrices finally used are as follows:
\begin{align*}
    \Qkf_Q &= \mathrm{diag} \left(\left[50 \;\; 1\ \;\; 10 \;\; 100\right]\right)\cdot10^{-20},\\
    \Qkf_P &= \mathrm{diag} \left(\left[10 \;\; 10\ \;\; 10 \;\; 10 \;\; 1 \;\; 1  \;\; 1 \;\; 1 \;\; 1\right]\right)\cdot10^{5},
\end{align*}
where $\mathrm{diag}(\cdot)$ is an operator that constructs a diagonal matrix with the elements of the argument vector on the main diagonal and zeros in the remaining elements. The covariance of the measurement noise has been kept at the same value ($\Rkf = \mathbf{I}_p \cdot 10^{-20}$).

\begin{figure*}
\centering
    \includegraphics[width=\linewidth]{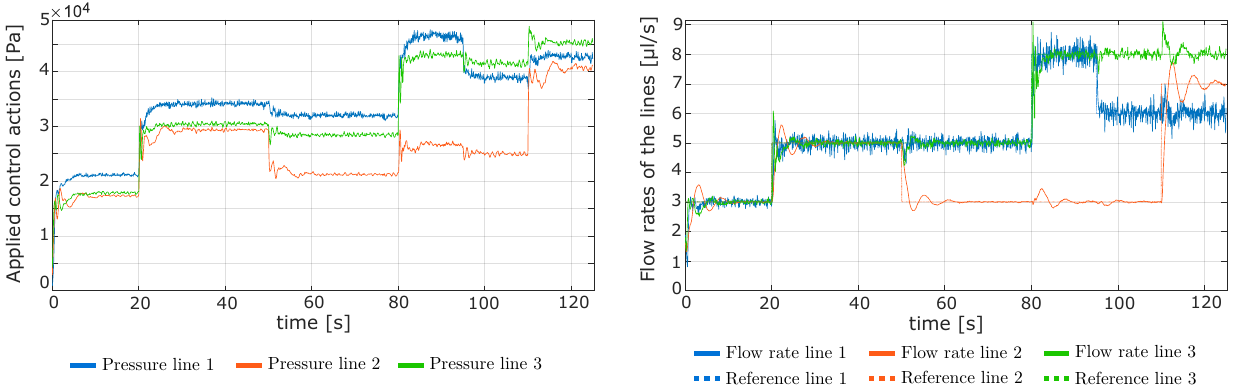}%
    \caption{Real operation of the complete MPC system. Pressures (left) and flow rates (right) in response to stepped setpoints, different for each line.}\label{fig:Exp_DIF1}
\end{figure*}

Fig.~\ref{fig:Exp_DIF1} shows the flow rates and pressures for the case already studied by simulation, with different stepped setpoints for each line. The results clearly show that the behavior of the middle line (line~2) is, as expected, different from the other two. In the flow rate plot of this case, as well as in those that will be presented later, it can be observed the previously mentioned oscillations in line~1 (in blue), which become more pronounced with higher flow rates. After several tests in the laboratory, it was concluded that this is due to unmodeled dynamics of this specific pressure regulator. There are other significant differences compared to the simulation, especially in the pressure plot, but it can still be seen that the proposed controller is able to achieve the desired flow rates in steady state. It can also be seen that disturbances appear in the lines when one of the others changes its value, but these are quickly corrected.

\begin{figure}[t]
\centering
  \includegraphics[width=\linewidth]{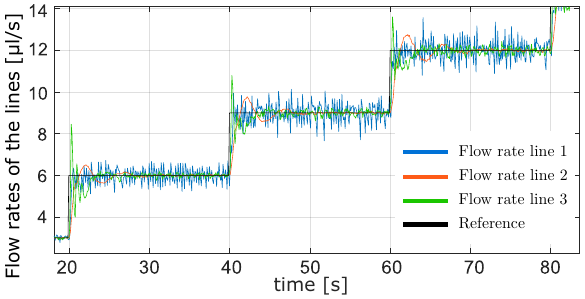}%
  \caption{Real operation of the complete MPC system. Response of the flow rates to equal step setpoints for all three~lines.}
  \label{fig:JuntoEscalonesPeq2}
\vspace{1.5\floatsep}
  \includegraphics[width=\linewidth]{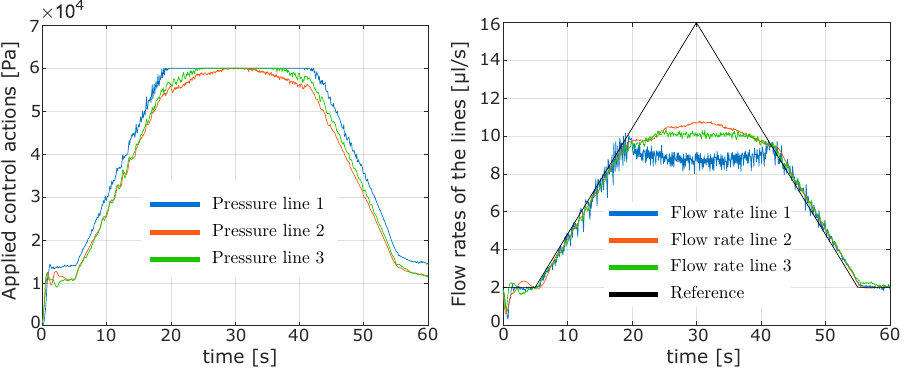}%
  \caption{Real operation of the complete MPC control system. Pressures (left) and flow rates (right) in response to equal triangle setpoints for all three lines. A maximum value of 60\,000~Pa has been imposed on the actions.}
  \label{fig:Exp_REST1}
\end{figure}

In addition to this test, several other experiments have been conducted. For example, Fig.~\ref{fig:JuntoEscalonesPeq2} shows the response of the system when the references are equal step functions for all three lines. Similar to the first case, it can be observed that the middle line takes longer to reach the reference than the other two. It is also noticeable that all three lines reach the steady state appropriately after the step functions. Some experiments have been done simply to illustrate compliance with the constraints imposed by the code. Fig.~\ref{fig:Exp_REST1} shows, for example, one of the experiments conducted, in which a constraint of 60\,000~Pa have been imposed on the maximum pressure in the pressure regulators. This pressure limit is established for verification purposes, as the normal operating limit is 150\,000~Pa. It can be seen that the flow rates follow the reference value up to the moment when the pressure limitation prevents reaching higher flow rates.

After the various experiments carried out on the real system, we can conclude that the designed controller allows to obtain flow rates at the chip inlet that adequately follow the desired reference values, while respecting the indicated constraints. The experiments also show that, although a simplification with a linear model has been used, the controller is capable of operating within the necessary working range, even in the presence of unexpected behaviors such as that of line~1. It has also been observed that modeling the dynamics of the pressure regulators is very important and that even two regulators of the same model and apparently identical can exhibit different behaviors.

\subsection{Comparison with a classical controller}

Finally, a comparison of the developed MPC controller with a classical controller has been performed by repeating the previous experiments but using an independent proportional-integral (PI) controller in parallel for each line. The possibility of including derivative action has been discarded due to the excessive presence of noise in the system. To make an objective comparison, the controllers have been optimized using the PID Tuner tool in Matlab, adjusting the proportional and integral gains, $k_p$ and $k_i$, to achieve the fastest response possible without overshoot. For lines 1 and 3, the gains are $k_p=5\cdot 10^{11}$~Pa/(m$^3$/s) and $k_i=2.5 \cdot 10^{12}$~Pa/m$^3$; for line~2, $k_p=8.5\cdot 10^{10}$~Pa/(m$^3$/s) and $k_i=1.5 \cdot 10^{12}$~Pa/m$^3$. The controllers include an anti-windup algorithm based on limiting the integral term.

\begin{figure*}
\centering
\vspace{\floatsep}
    \includegraphics[width=0.98\linewidth]{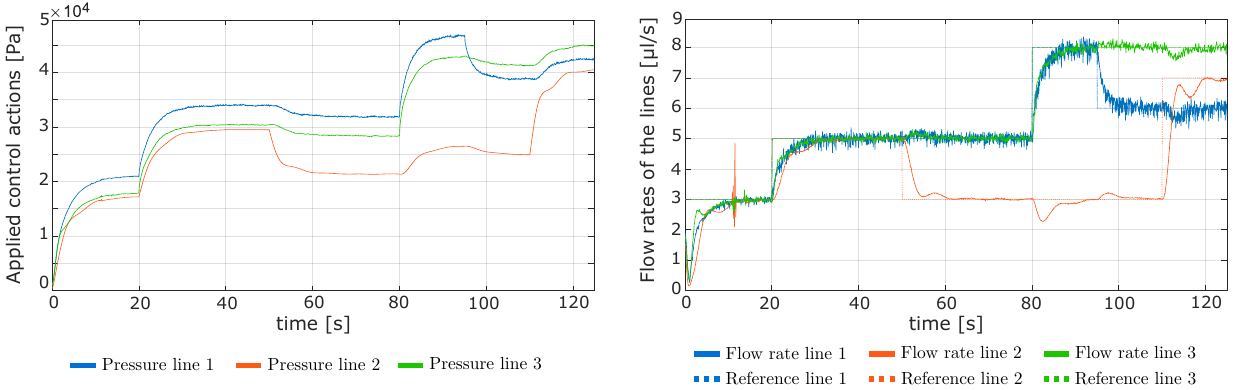}\\
    \caption{Real operation of the classical control system. Pressures (left) and flow rates (right) in response to different step references for each line.}\label{fig:JuntoDIF1_PID}
\end{figure*}

\begin{figure}[p]
\centering
  \includegraphics[width=\linewidth]{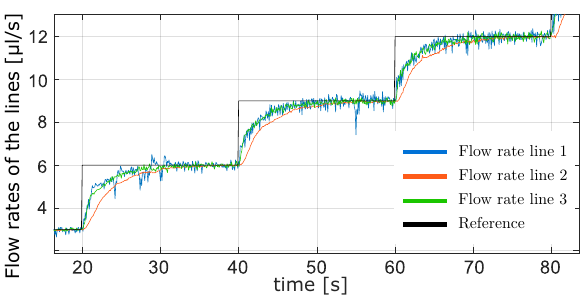}\\
  \caption{Real operation of the classical control system. Response of the flow rates to equal step setpoints for all three~lines.}
  \label{fig:JuntoEscalonesPeq2PI}
\vspace{1.5\floatsep}
  \includegraphics[width=\linewidth]{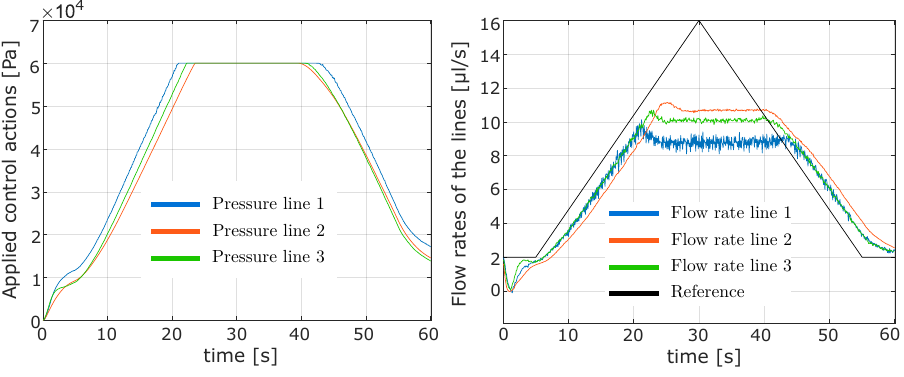}\\
  \caption{Real operation of the classical control system. Pressures (left) and flow rates (right) in response to equal triangle references for all three lines. A maximum value of 60\,000~Pa has been imposed on the actions.}
  \label{fig:Exp_REST1_PI}
\end{figure}

In Figs.~\ref{fig:JuntoDIF1_PID} and \ref{fig:JuntoEscalonesPeq2PI}, the results obtained for the two stepped setpoint references are presented. It can be observed that the response of the system with the PI controllers is slower than that obtained in equivalent tests with the MPC, both in tracking the reference and in correcting one line when changes occur in the others. Specifically, the response time with the classical controller is between three and five times longer than with the MPC. Fig.~\ref{fig:Exp_REST1_PI} shows the response to the triangular reference signal, including the aforementioned limitation of 60\,000~Pa on the maximum control action. It can be seen that the flow rates do not exactly follow the reference even before reaching saturation, contrary to what happened with the MPC (see Fig.~\ref{fig:Exp_REST1}). In this test, the MPC controller is also faster when the pressure saturation ends and returns to tracking the descending desired value. Finally, Table \ref{tab:comparativaPIMPC} shows the tracking errors for each of the flow rates for the three analyzed reference signals using each of the controllers. It can be seen that the MPC controller achieves lower errors in all cases.

\renewcommand{\arraystretch}{1.3} 
\begin{table}[p]
    \centering
    \vspace{-2em}
    \caption{Comparison of results of MPC and classical controller. Root Mean Square Errors (RMSE) in $\mu$l/s.}
    \begin{tabular}{ccccc}
        \hline
        Reference && RMSE $Q_1$ & RMSE $Q_2$ & RMSE $Q_3$\\
        \hline
         Different & MPC & {0,325} & 0,384 & 0,262 \\
         steps & PI & 0,490 & 0,705 & 0,425 \\
         \hline
        Equal & MPC & 0,566 & 0,475 &0,411 \\
        steps & PI & 0,723 & 0,980 & 0,681 \\
         \hline         
         \multirow{2}{*}{Triangle}& MPC & 2,743 & 1,862 & 2,065 \\
         & PI & 2,857 & 2,218 & 2,200 \\
         \hline
    \end{tabular}
    \label{tab:comparativaPIMPC}
\end{table}
\renewcommand{\arraystretch}{1} 

\section{Conclusions}
\label{sec:conclusiones}

In this article, the design of a predictive controller for a microfluidic system has been described. Initially, it was necessary to develop a model to describe the fluid dynamics through the different channels and lines, as well as the behavior of the low-level sensors and regulators present in the system. Based on this, a state observer and the controller itself were designed, which were subsequently validated both by simulation and in the real system. The presented results show that the controller is effective in independently controlling the three flow rates circulating inside the microfluidic chip, even in the presence of noise and certain simplifications and uncertainties inherent in the model, and that the results improve upon what could be achieved with three independent classical controllers. It has also been observed that small variations in the operating conditions, e.g., the deformation of one of the conduits in the lines, can have a significant impact on the dynamics of the system. Thus, future research will focus on the development of online parameter estimation algorithms so that the control system can adapt to these changes.

\section*{Acknowledgments}

This work has been partially funded by the Government of Aragon, through grants aimed at promoting research activity of research groups (groups T45\_23R and T73\_23R), by the European Regional Development Fund (ERDF) and by the ``Programa de Becas y Ayudas'' of the Instituto de Investigación en Ingeniería de Aragón (I3A).

The authors would like to thank the "DIseño, desArrollo y Modelado cOmputacional de Materiales avanzaDos (DIAMOMD)" research group of ITAINNOVA, whose test bench has enabled the experiments presented in this work.

\bibliographystyle{elsarticle-harv}

\end{document}